# Tetragonality induced superconductivity in anti-ThCr$_2$Si$_2$-type $RE_2O_2Bi$ ($RE$ = rare earth) with Bi square net

Ryosuke Sei,[a,b] Hideyuki Kawasoko,[a] Kota Matsumoto,[a] Masato Arimitsu,[a] Kyohei Terakado,[a] Daichi Oka,[a] Shintaro Fukuda,[a] Noriaki Kimura,[c] Hidetaka Kasai,[d] Eiji Nishibori,[d] Kenji Ohoyama,[e] Akinori Hoshikawa,[f] Toru Ishigaki,[f] Tetsuya Hasegawa,[b] and Tomoteru Fukumura*[a,g]

We report a series of layered superconductors, anti-ThCr$_2$Si$_2$-type $RE_2O_2Bi$ ($RE$ = rare earth), composed of electrically conductive Bi square nets and magnetic insulating $RE_2O_2$ layers. The superconductivity was induced by separating Bi square nets as a result of excess oxygen incorporation, irrespective of the presence of magnetic ordering in $RE_2O_2$ layers. Intriguingly, the transition temperature of all $RE_2O_2Bi$ including nonmagnetic Y$_2$O$_2$Bi was approximately scaled by the unit cell tetragonality ($c/a$), implying a key role of relative separation of the Bi square nets to induce the superconductivity.

## Introduction

Layered compounds have demonstrated intriguing superconducting states such as high temperature superconductivity, heavy fermion superconductivity, spin-triplet superconductivity, and topological superconductivity.[1–6] Such states were often emerged by chemical substitution and intercalation.[7–9] For example, high temperature superconductivity and topological superconductivity were achieved by carrier doping via aliovalent substitution in (La,Ba)$_2$CuO$_4$ and metal intercalation in Cu$_x$Bi$_2$Se$_3$, respectively.[2,6] Also, chemical pressure effect by isovalent substitution increased superconducting transition temperature in $RE$Ni$_2$B$_2$C ($RE$ = rare earth) and BaFe$_2$(As,P)$_2$.[10,11] In case of the intercalation, large interlayer expansion in $M$NCl ($M$ = Ti, Zr, and Hf) and FeSe e.g. by organic molecules is crucial to enhance superconducting transition temperature, indicating a principal role of their two-dimensional electronic states.[12,13]

Anti-ThCr$_2$Si$_2$-type $RE_2O_2Bi$ is composed of alternating stack of insulating $RE_2O_2$ layers and electrically conducting Bi square net with Bi$^{2-}$ valence (Fig. 1).[14,15] A series of $RE_2O_2Bi$ was reported to show metal-insulator transition driven by increased Bi–Bi interatomic distance in each Bi square net through different $RE$ ions (i.e. increased $a$-axis length).[14] Recently, Y$_2$O$_2$Bi and Er$_2$O$_2$Bi were found to be superconducting near 2 K by slight increase in interlayer distance between Bi square nets via excess oxygen incorporation (i.e. increased $c$ parameter) without carrier doping.[16,17] In this study, we report a series of new layered superconductors $RE_2O_2Bi$, that underwent superconducting transition around or below 2 K. The superconducting transition temperature ($T_c$) was not varied by presence or absence of the antiferromagnetic ordering in $RE_2O_2Bi$, but was universally scaled by the unit cell tetragonality $c/a$: that is a key parameter for emergence of the superconductivity.

## Experimental

Stoichiometric and excess-oxygen-incorporated $RE_2O_2Bi$ ($RE$ = Tb, Dy, Y, Er, and Lu) polycrystals were synthesized by solid-state reaction. Tb$_4$O$_7$ (99.9%), Dy$_2$O$_3$ (99.9%), Y$_2$O$_3$ (99.9%), Er$_2$O$_3$ (99.9%), and Lu$_2$O$_3$ (99.9%) powders were heated at 1000 °C in furnace for 10 hours to remove moisture. CaCO$_3$ (99.9%) powder was heated at 1000 °C in furnace for 10 hours to decompose into CaO, which serves as an oxidant to synthesize excess-oxygen-incorporated $RE_2O_2Bi$.[17] Tb (99.9%), Tb$_4$O$_7$ (99.9%), Dy (99.9%), Dy$_2$O$_3$ (99.9%), Y (99.9%), Y$_2$O$_3$ (99.9%), Lu (99.9%), Lu$_2$O$_3$ (99.9%), Bi (99.9%), and CaO powders were mixed for various nominal compositions (Tables S1–S3) and pelletized under 20 MPa in nitrogen-filled glove box. The pellets covered with Ta foils were sintered in evacuated quartz tubes at 500 °C for 7.5 hours, followed by sintering at 1000 °C for 20 hours. The sintered products were ground and pelletized under 30 MPa again in the glove box, and then the pellets covered with Ta foils were sintered in evacuated quartz tubes

[a.] Department of Chemistry, Graduate School of Science, Tohoku University, Sendai 980-8578, Japan.
E-mail: tomoteru.fukumura.e4@tohoku.ac.jp
[b.] Department of Chemistry, Graduate School of Science, The University of Tokyo, Tokyo 113-0033, Japan.
[c.] Department of Physics, Graduate School of Science; Center for Low Temperature Science, Tohoku University, Sendai 980-8578, Japan.
[d.] Division of Physics and Tsukuba Research Center for Energy Materials Science, Faculty of Pure and Applied Sciences, University of Tsukuba, Tsukuba, 305-8571, Japan.
[e.] Graduate School of Science and Engineering, Ibaraki University, Tokai 319-1106, Japan.
[f.] Frontier Research Center for Applied Atomic Sciences, Ibaraki University, Tokai, 319-1106, Japan.
[g.] Advanced Institute for Materials Research and Core Research Cluster, Tohoku University, Sendai, 980-8577, Japan.

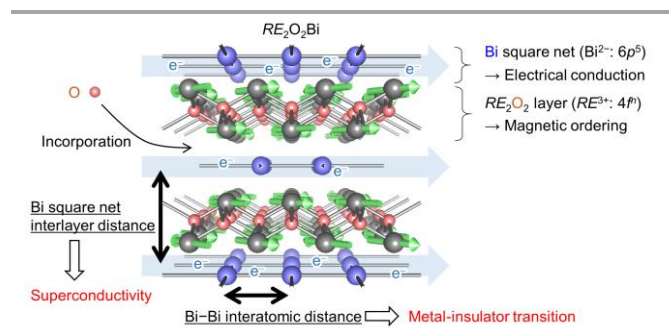

**Fig. 1** Schematic crystal structure of $RE_2O_2Bi$ ($RE$ = rare earth). In $RE_2O_2Bi$, Bi square nets and $RE_2O_2$ layers are responsible for electrical conduction and magnetic ordering, respectively. Incorporation of excess oxygen expands interlayer distance of Bi square nets, inducing superconductivity in $RE_2O_2Bi$.

at 1000 °C for 10 hours. CaO powder was used as an oxidant for the incorporation of excess oxygen in $RE_2O_2Bi$ ($RE$ = Tb, Dy, and Lu), and Ca was evaporated or precipitated after synthesis without incorporating into $RE_2O_2Bi$ (Fig. S1). In $Y_2O_2Bi$, the obtained pellet covered with Ta foils were sintered again in evacuated quartz tubes containing $Y_2O_3$ pellet at 1000 °C for 10 hours to incorporate excess oxygen (Table S4). Crystal structures were evaluated by powder X-ray diffraction (XRD with Cu K$\alpha$ radiation; D8 DISCOVER, Bruker AXS), synchrotron X-ray powder diffraction (SPring-8 BL02B2, wavelength: 0.44 Å, exposure time: 90 min), and powder neutron diffraction (J-PARC, BL20, iMATERIA diffractometer) at room temperature, 100 K, and low temperatures, respectively. Rietveld analysis was performed by using RIETAN-FP (Ref. 18), Synchrotron Powder (SP) (Ref. 19), and FullProf (Ref. 20) to identify the crystal phases and their lattice constants (see Figs. S2–S4 and Tables S1–S4). It is noted that rather low purity of $RE_2O_2Bi$ phase in Table S2 would be caused by decomposition of $RE_2O_2Bi$ into non-superconducting $RE_2O_3$ and Bi promoted by excess oxygen incorporation, where the lifetime was within few days in air. These factors significantly influenced the results of XRD measured in air. On the other hand, their influence on the values of $T_c$ was negligible, because of the non-superconducting impurity phases ($RE_2O_3$ and Bi) and the inert He-gas atmosphere for transport measurements. Surface morphology and chemical composition were investigated by scanning electron microscope equipped with energy dispersive X-ray spectroscopy (SEM-EDX; JEOL Ltd., JSM-7100F). Transport properties were evaluated by a standard four terminal method in physical properties measurement system (PPMS, Quantum Design) and dilution refrigerator (Kelvinox TLM, Oxford Instruments). The crystal structure was drawn with the VESTA.[21]

## Results and discussion

The $a$- and $c$- axis lengths of all the $RE_2O_2Bi$ ($RE$ = Tb, Dy, Y, Er, and Lu) with various amounts of oxygen are shown in Fig. 2. With increasing ionic radii of $RE^{3+}$ ions, both $a$- and $c$- axis lengths increased monotonically for stoichiometric $RE_2O_2Bi$ (open symbols in Fig. 2), being consistent with the previous study.[14] The anti-ThCr$_2$Si$_2$-type structure of $Y_2O_2Bi$ was also confirmed to be preserved regardless of excess oxygen incorporation from synchrotron X-ray diffraction (Fig. S5). It is noted that $RE$−O bond lengths indicate that $RE$ ion was trivalent (Tables S1–4). Irrespective of the Bi–Bi interatomic distance (i.e. $a$-axis length), all the stoichiometric $RE_2O_2Bi$ showed metallic conduction without superconducting transition down to around 2.0 K (Fig. 3). The resistivity anomaly of $Tb_2O_2Bi$ for 10–40 K corresponds to phase transitions including antiferromagnetic ordering (Fig. S6).

Fig. 4a shows temperature dependence of resistivity for stoichiometric and excess-oxygen-incorporated $RE_2O_2Bi$. For the stoichiometric $RE_2O_2Bi$ (dashed curve), $Y_2O_2Bi$, $Er_2O_2Bi$, and $Lu_2O_2Bi$ became superconducting at 1.50 K, 1.31 K, and 1.26 K, respectively, while $Tb_2O_2Bi$ and $Dy_2O_2Bi$ were not superconducting down to 55 mK. The absence of superconductivity in $Tb_2O_2Bi$ and $Dy_2O_2Bi$ with the longer $c$-axis lengths than those of superconducting $Y_2O_2Bi$, $Er_2O_2Bi$, and $Lu_2O_2Bi$ means that the absolute value of $c$-axis length is not crucial for the emergence of superconductivity.

For the excess-oxygen-incorporated $RE_2O_2Bi$, whose $c$-axis lengths were longer while $a$-axis lengths were almost constant (solid symbols in Fig. 2), all the $RE_2O_2Bi$ showed superconductivity at 1.90–2.29 K (solid curve in Fig. 4a), irrespective of the antiferromagnetic ground state in the stoichiometric $RE_2O_2Bi$ except for $Y_2O_2Bi$,[14,22] in contrast with considerably reduced $T_c$ in superconductor/antiferromagnet bilayer systems.[23–25] The superconducting transition temperatures increased with increasing $c$-axis length in each

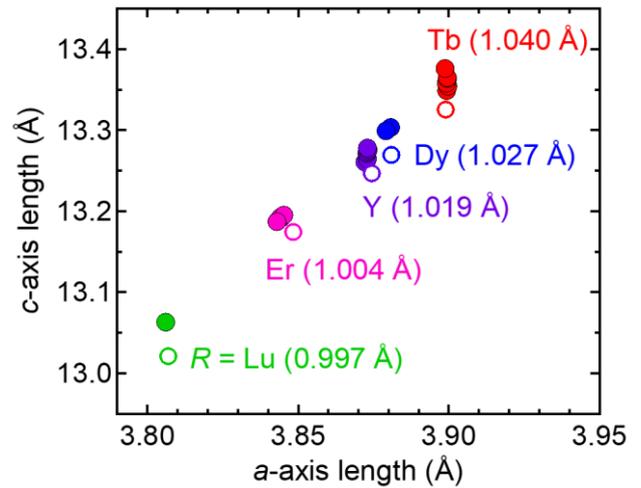

**Fig. 2** $a$- and $c$- axis lengths of $RE_2O_2Bi$ ($RE$ = Tb, Dy, Y, Er, and Lu) evaluated from Rietveld analysis of powder X-ray diffraction patterns. Open and solid circles correspond to stoichiometric and excess-oxygen-incorporated $RE_2O_2Bi$ samples, respectively. Numeric values in brackets are ionic radii of $RE^{3+}$ ions.[30]

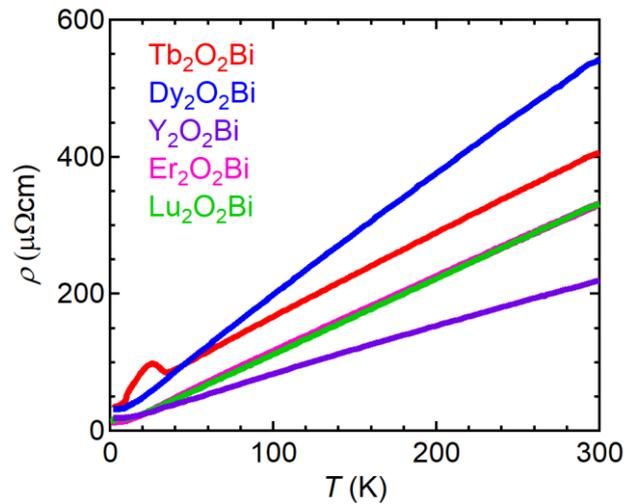

**Fig. 3** Temperature dependence of resistivity for stoichiometric $RE_2O_2Bi$ ($RE$ = Tb, Dy, Y, Er, and Lu). Data of $Y_2O_2Bi$ and $Er_2O_2Bi$ were cited from Refs. 16 and 17, respectively.

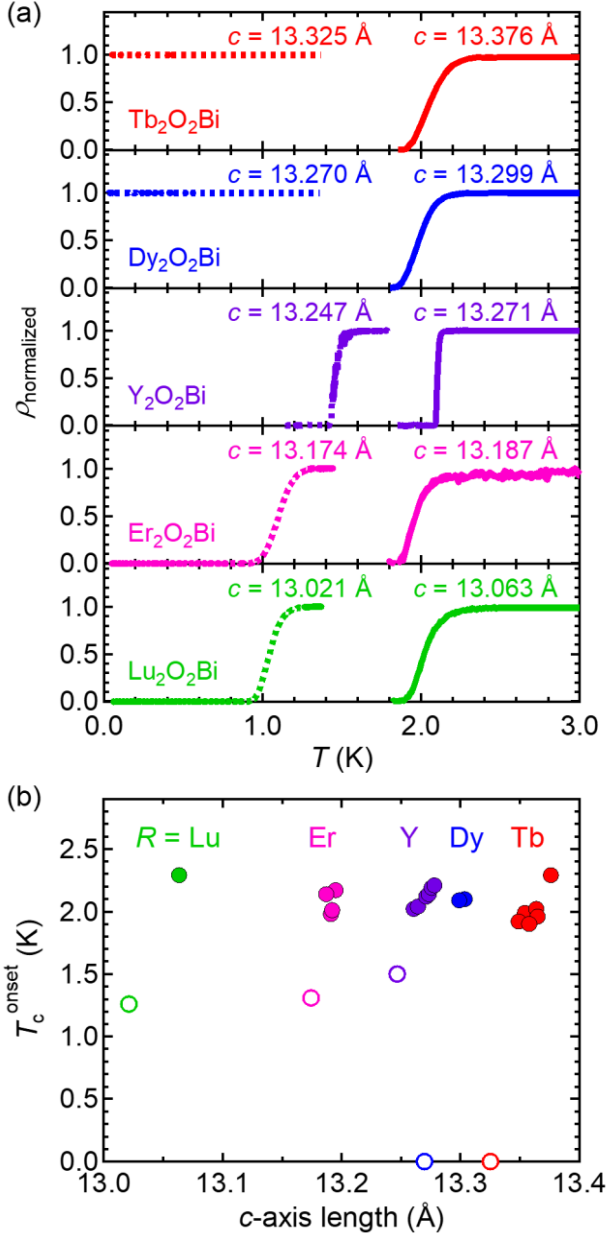

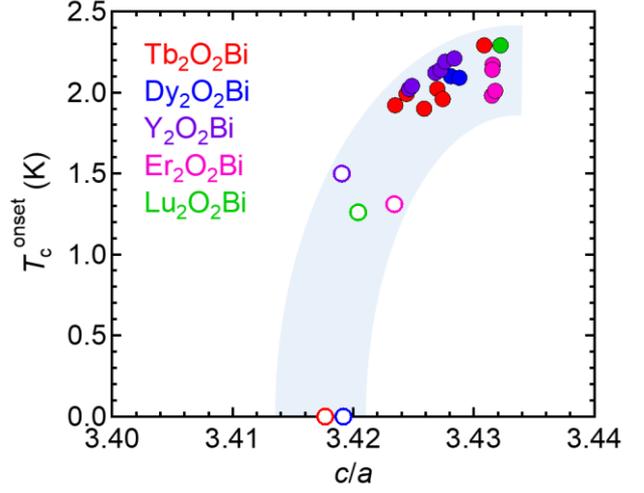

**Fig. 5** Relationship between tetragonality $c/a$ and superconducting transition temperature for $RE_2O_2Bi$ ($RE$ = Tb, Dy, Y, Er, and Lu). Open and solid circles correspond to stoichiometric and excess-oxygen-incorporated $RE_2O_2Bi$ samples, respectively.

**Fig. 4** (a) Temperature dependence of normalized resistivity for stoichiometric (dashed curve) and excess-oxygen-incorporated (solid curve) $RE_2O_2Bi$ ($RE$ = Tb, Dy, Y, Er, and Lu) below 3 K. (b) Relationship between c-axis length and superconducting transition temperature for $RE_2O_2Bi$ ($RE$ = Tb, Dy, Y, Er, and Lu). Open and solid circles correspond to stoichiometric and excess-oxygen-incorporated $RE_2O_2Bi$ samples, respectively. Data of excess-oxygen-incorporated $Y_2O_2Bi$ and $Er_2O_2Bi$ were cited from Refs. 16 and 17, respectively.

$RE_2O_2Bi$ (Fig. 4b). Since the stoichiometric $RE_2O_2Bi$ have a metallic ground state, influence of carrier doping by excess oxygen was insignificant for the emergence of superconductivity, as was previously discussed in $Y_2O_2Bi$.[16] Accordingly, the expansion of c-axis length corresponding to the separation of Bi square nets was responsible for superconductivity, while the threshold values of c-axis lengths were different for each $RE_2O_2Bi$. The broad superconducting transition in $RE_2O_2Bi$ other than $Y_2O_2Bi$ might be caused by the low purity and/or the magnetic ordering.

Intriguingly, the c-axis length normalized by the a-axis length showed universal trend for the emergence of superconductivity, in which the increased tetragonality ($c/a$) corresponds to tetragonal elongation of the unit cell. As shown in Fig. 5, the superconductivity emerged above the tetragonality of ~3.42; $T_c$ increased steeply and saturated to be around 2 K with increasing the tetragonality, suggesting intimate relation between the tetragonality and the superconductivity. A linear correlation between the tetragonality and $T_c$ was previously observed in heavy fermion superconductors $CeMIn_5$ and $PuMGa_5$ ($M$ = Co, Rh, and Ir), in which the superconductivity was attributed to the relevant spin fluctuations.[26,27] In case of $RE_2O_2Bi$, however, the $T_c$ was not influenced by presence or absence of magnetic ordering. For instance, antiferromagnetic $Tb_2O_2Bi$ and non-magnetic $Y_2O_2Bi$ showed similar $T_c$ around 2 K by excess oxygen incorporation, probably ruling out the spin fluctuation scenario in $RE_2O_2Bi$ system. On the other hand, the increased tetragonality in $RE_2O_2Bi$ might modify the shape of Fermi surface, possibly contributing to emergence of superconductivity, because the correlation between interlayer distance and $T_c$ was observed in $ThCr_2Si_2$-type $ANi_2Pn_2$ ($A$ = Ca, Sr, Ba; $Pn$ = P, As).[28,29] Also, the tetragonality dependence of $T_c$ observed in $RE_2O_2Bi$ implies that both the two-dimensionality of Bi square nets enhanced by longer c-axis length and the phonon frequency of Bi–Bi bonds increased by shorter a-axis length were beneficial for higher $T_c$. Thus, further investigation of electronic and phonon band structures is needed.

## Conclusions

The separation of Bi square nets was essential for the emergence of superconductivity in Bi square net superconductors $RE_2O_2Bi$. Interestingly, the tetragonality $c/a$ in $RE_2O_2Bi$ was found to govern the $T_c$ irrespective of presence or absence of magnetic ordering, raising a possibility that the tetragonality is the key parameter for other square net layered superconductors. In addition, competing phase between antiferromagnetism and superconductivity as a function of separation of Bi square nets suggests that those Bi square net layered crystals are rich playground to investigate interplay of superconductivity and magnetic ordering.

## Conflicts of interest

There are no conflicts to declare.

## Acknowledgements

This study was supported by JSPS KAKENHI (26105002, 16H06441), JST-CREST, and Yazaki Memorial Foundation for Science and Technology. The synchrotron radiation experiments were performed at the BL02B2 of SPring-8 with the approval of the Japan Synchrotron Radiation Research Institute (JASRI) (Proposal No. 2017A0074 and 2019A0068). The neutron experiments at the Materials and Life Science Experimental Facility of the J-PARC was performed under a user program (Proposal No. 2017A0117).

## Notes and references